\pdfoutput=1

\documentclass[preprint,authoryear,12pt]{elsarticle}

\usepackage{graphicx}
\usepackage{amssymb}

\journal{Advances in Space Research}

\begin{document}


\begin{frontmatter}

\title{Anticorrelated temperature--density profiles in \\
       the quiet solar corona and coronal mass ejections: \\
       Approach based on the spin-type Hamiltonians}

\author[GAISh,IKI]{Yurii V. Dumin\corref{cor}}
\cortext[cor]{Corresponding author}%
\ead{dumin@yahoo.com}

\address[GAISh]{P.K.~Sternberg Astronomical Institute
                of M.V.~Lomonosov Moscow State University,\\
                Universitetskii prosp.\ 13, 119234, Moscow, Russia}

\address[IKI]{Space Research Institute
              of Russian Academy of Sciences,\\
              Profsoyuznaya str.\ 84/32, 117997, Moscow, Russia}

\begin{abstract}
The problem of solar corona heating remains one of key puzzles in astrophysics
for a few decades; but none of the proposed mechanisms can give a definitive
answer to this question.
As a result, the novel scenarios are still suggested.
Here, we perform a critical consideration of the recently-proposed mechanism
for the formation of anticorrelated temperature and density profiles due to
specific features of relaxation in the strongly non-equilibrium plasmas
described by the so-called spin-type Hamiltonians (L.~Casetti \& S.~Gupta.
\textit{Eur.\ Phys.\ J.\ B} 87, 91, 2014; T.N.~Teles \textit{et al.}
\textit{Phys.\ Rev.\ E} 92, 020101(R), 2015).
We employ the universal property of these systems to produce the long-lived
anticorrelated temperature--density distributions and analyze their most
important qualitative features in the context of coronal plasmas.
As follows from our consideration, such anticorrelated profiles can be hardly
relevant to explanation of the temperature distribution in the quiet solar
corona.
However, they might be interesting for the interpretation of the large-scale
inhomogeneity in the coronal mass ejections and the resulting solar
wind.
\end{abstract}


\end{frontmatter}

\parindent=0.5 cm


\section{Introduction}
\label{sec:Intro}

The mechanism of heating the solar corona, \textit{i.e.}, a sharp increase of
temperature in its upper, very rarefied layers, remains one of key problems
in astrophysics since the middle of the last century.
Despite a lot of the proposed scenarios---based either on wave absorption
or micro- and nanoflares---none of them can provide an ultimate resolution
of the heating problem
\citep[\textit{e.g.}, reviews][and references therein]{wal03,erd07}.
As a result, the new original hypotheses continue to be suggested up to the
present time.

One of them is an attempt to explain the coronal heating by the specific
properties of the model plasmas described by the so-called spin-type
Hamiltonians, whose mathematical theory is developed for almost a century.
As regards the equilibrium parameters of these systems, the most of them can
be calculated analytically, at least, in one- and two-dimensional cases
\citep{isi71,rum80}; but they are interesting mostly in the condensed-matter
physics.
Studying the non-equilibrium behaviour of such Hamiltonians began much later,
and it is usually performed by a combination of analytical and numerical
methods \citep{cas14,tel15}.
It is important that the scope of applicability of the corresponding results
becomes much wider and, in particular, they may be interesting also for
the physics of plasmas and gravitating systems.%
\footnote{
There were also some earlier applications of the spin-type models to studying
the strongly non-equilibrium processes in the continuous media, \textit{e.g.},
in our work \citep{dum09}; but they referred to the so-called $ {\varphi}^4 $
nonlinear field model rather than to the plasma physics.}

It should be mentioned that a possible application of the specific features
of relaxation in such plasmas to the problem of solar corona heating was
discussed at a number of conferences on the statistical physics of strongly
non-equilibrium systems for more than a decade.%
\footnote{
In particular, this was a series of workshops in the Max Planck Institute for
the Physics of Complex Systems (Dresden, Germany):
``Fluctuation and Dissipation Phenomena in Driven Systems far from
Equilibrium'' (16--18.07.2007),
``Many-Body Systems far from Equilibrium: Fluctuations, Slow Dynamics and
Long-Range Interactions'' (16--27.02.2009),
``Large Fluctuations in Non-Equilibrium Systems'' (04--15.07.2011),
``Small Systems far from Equilibrium: Order, Correlations, and Fluctuations''
(14--18.10.2013),
``Stochastic Thermodynamics: Experiment and Theory'' (10--14.09.2018).}
Unfortunately, the major part of their participants were specialists in
theoretical and mathematical physics, unrelated immediately to
the astrophysical or plasma-physics research.
In the last years, a few papers on this subject went out of press in
the journals on statistical and mathematical physics
\citep[\textit{e.g.}, the above-cited works][]{cas14,tel15} but, again,
they remained almost unknown to the wide astronomical community.

Therefore, it is one of the aims of the present article to draw attention
of astronomers and astrophysicists to the respective ideas.
In Sec.~\ref{sec:Model} we give a brief overview of the corresponding
mathematical formalism and in Sec.~\ref{sec:Corona} apply it to the coronal
plasmas.

\section{Dynamics of the strongly non-equilibrium plasmas}
\label{sec:Model}

\subsection{Spin-type models}
\label{sec:Spin-type}

Let us begin with a brief qualitative review how the plasmas can be described
by the spin-type Hamiltonians and what are the main features of relaxation of
the non-equilibrium states in such models; more mathematical details can be
found in the papers cited in Introduction.
The main idea is---by using certain approximations---to reduce a Hamiltonian
of the system of charged particles (\textit{i.e.}, plasma) to the Hamiltonian
of a spin system.
Next, since the theory of spin-type Hamiltonians is an extensive and
well-developed branch of mathematical physics, involving a number of nontrivial
and quite universal findings, it becomes possible to transfer the corresponding
results to the plasma phenomena.

\begin{figure}
\begin{center}
\includegraphics[width=0.9\hsize]{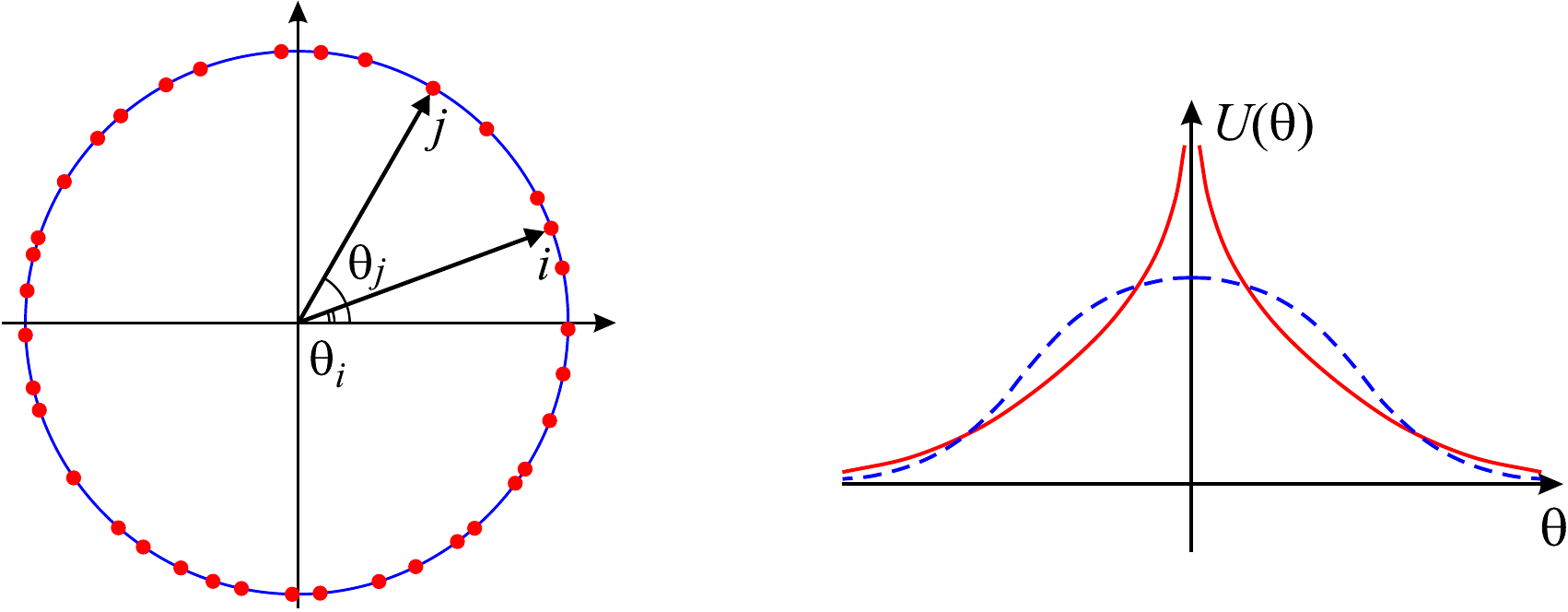}
\end{center}
\caption{Sketch of the plasma description by the spin-type model:
(left) geometric configuration of the one-dimensional system and
(right) approximation of the Coulomb's interparticle potential
(red solid curve) by the lowest-order Fourier harmonic (blue
dashed curve).}
\label{fig:Model}
\end{figure}

We shall consider the simplest one-dimensional system, which is a reasonable
approximation for the strongly-magnetized plasmas.
For simplicity, we assume the periodic boundary conditions, \textit{i.e.},
the region under consideration is topologically equivalent to a circle.
Then, from the mathematical point of view, positions of the charged particles
can be conveniently characterized by the angles~$ {\theta}_i $, defined with
respect to the center of the circle; see left panel in Fig.~\ref{fig:Model}.
Next, we approximate the Coulomb's potential~$ U(\theta) $ just by
the lowest-order Fourier harmonic, namely, cosine function with a period
corresponding to the size of the system; see right panel in the same figure.%
\footnote{
Strictly speaking, because of a singularity of the Coulomb's potential
in zero, the Fourier expansion is ill defined.
So, some artificial Fourier-type approximation is actually employed here.}

As a result, the ionic Hamiltonian takes the form:
\begin{equation}
{\cal H} = \, \frac{1}{2} \sum\limits_{i=1}^N p_i^2 +
           \frac{J}{N} \sum\limits_{i=1}^N \sum\limits_{j=1}^{i-1}
           \big[ 1 - \cos({\theta}_i - {\theta}_j)\big] \, ,
\label{eq:Hamiltonian}
\end{equation}
where
$ p_i $~are the momenta of ions (which, for simplicity, are assumed to be of
a unitary mass),
$ N $~is their total number, and
$ J $~is the interaction parameter commonly introduced in the spin-type models,
which can be expressed though the parameters of the Coulomb's system.
Since the entire plasma must be electrically neutral, the above-mentioned
ionic system should be embedded into the uniform electron gas, which
corresponds to the well-known OCP (one-component plasma) approximation.

Let us also mention that in the case of plasma (where particles of the same
sign repel each other) the interaction parameter~$ J $ in
formula~(\ref{eq:Hamiltonian}) is negative.
By terminology used in the condensed-matter physics, this is called the
``antiferromagnetic'' type of interaction.
At positive~$ J $ (``ferromagnetic'' interaction), the particles are attracted
to each other; this is an analogue of the gravitating system.

\begin{figure}
\begin{center}
\includegraphics[width=1.0\hsize]{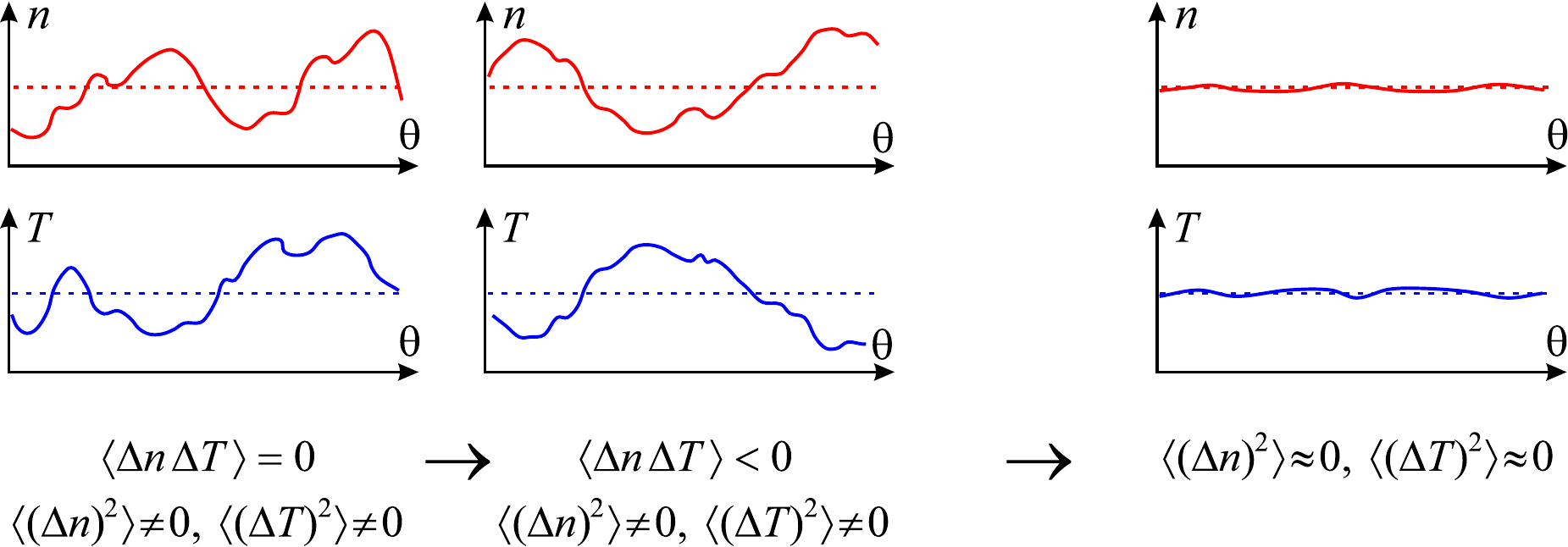}
\end{center}
\caption{Sketch of the two stages of relaxation in the strongly
non-equilibrium plasmas predicted by the spin-type models.
The empty space between the second and third plots implies
a much longer duration of the second stage as compared to
the first stage of the relaxation.
\label{fig:Relaxation}}
\end{figure}

Therefore, the original Coulomb's system is reduced to the spin-type
system~(\ref{eq:Hamiltonian}), where the cosine-type interaction between
the spins is a counterpart of the Coulomb's interaction between the charged
particles.%
\footnote{
An additional constant term in the sum was introduced in order to get
an exact correspondence to the spin interaction.
This term evidently does not affect the equations of motion.}
Both equilibrium and non-equilibrium properties of such Hamiltonians were
investigated in very much detail in the physics of condensed matter.
In the context of plasma physics, these Hamiltonians were studied in
the papers by \citet{cas14,tel15}.
As a result, the following most important properties were established:
\begin{enumerate}
\item
If plasma was originally created in a strongly non-equilibrium state
(\textit{i.e.}, experienced considerable fluctuations both in the temperature
and density), then its relaxation proceeds in two stages:
Firstly, some quasi-equilibrium state is quickly formed.
Then, at a much longer temporal scale, this state relaxes to the genuine
thermodynamic equilibrium.
\item
A universal property of the above-mentioned quasi-equilibrium state is
anti-correlation between the spatial inhomogeneities of temperature and
density.
\end{enumerate}
This peculiarities of the relaxation are schematically illustrated in
Fig.~\ref{fig:Relaxation}.

\begin{figure}[t]
\begin{center}
\includegraphics[width=0.85\hsize]{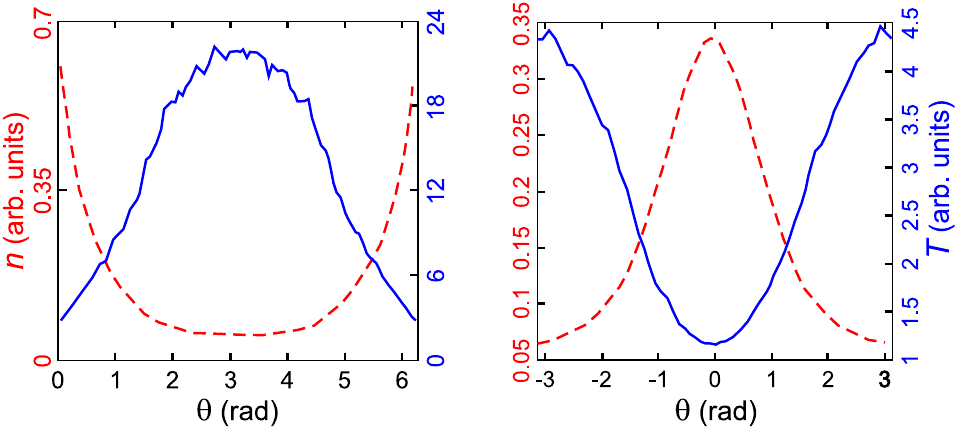}
\end{center}
\caption{Examples of the anticorrelated temperature (blue solid curves)
and density (red dashed curves) computed for plasmas described by
the spin-type Hamiltonians:
(left panel) reprinted Fig.~2 by \citet{cas14}
with permission from Springer Nature Switzerland AG, {\copyright}~2014
and
(right panel) reprinted Fig.~2 by \citet{tel15}
with permission from the American Physical Society, {\copyright}~2015.
Dimensionless units, employed in the original simulations,
are used in all plots.}
\label{fig:Simulations}
\end{figure}

Examples of the particular calculations are given in
Fig.~\ref{fig:Simulations}.
Left panel corresponds to the case of ``antiferromagnetic'' (Coulomb's)
interparticle interaction with imposition of the permanent external field
and non-Maxwellian initial velocity distribution.
Right panel corresponds to the ``ferromagnetic'' (gravitational) type of
interaction with a Maxwellian initial velocity distribution disturbed by
a short sudden pulse of the external field.
It is seen that, despite the very different simulation setups, the final
results are qualitatively the same.
So, formation of the anticorrelated profiles is a generic property of
the systems described by the Hamiltonian~(\ref{eq:Hamiltonian}).

An alternative explanation of the same phenomenon, also proposed by
\citet{tel15}, is based on the well-known effect of Landau damping
\citep[\textit{e.g.},][]{lif81,som12}.
Namely, if the original plasma was strongly non-equilibrium (turbulent),
then it should involve a considerable amount of the stochastic waves.
They will subsequently experience the Landau damping and, thereby, form
a non-Maxwellian tail of the velocity distribution function.
Then, the so-called effect of ``velocity filtration'' comes into play,
when only the most energetic particles can reach the regions of reduced
density, corresponding to the higher potential energy.%
\footnote{
To avoid misunderstanding, let us emphasize that the velocity filtration
mechanism was used by the above-cited authors merely as one of pictorial
ways for the interpretation of the results of calculations.
It is not employed in the self-consistent treatment of the spin-type
Hamiltonians.}
As a result, the anticorrelated temperature and density profiles should be
again formed.

\subsection{Physical interpretation}
\label{sec:Physical}

From our point of view, the results outlined in Sec.~\ref{sec:Spin-type}
can be interpreted even in a more pictorial way, which is actually independent
of the specific features of the spin-type Hamiltonians:
\begin{enumerate}
\item
Existence of the first (fast) and second (slow) stages in the dynamics of
strongly non-equilibrium plasmas immediately follows from the two very
different time scales for the relaxation of pressure,~$ ({\Delta}t)_p $,
on the one hand, and for the density and temperature, $ ({\Delta}t)_n $ and
$ ({\Delta}t)_T $, on the other hand.
Really, the characteristic rate of relaxation of the pressure is about
the speed of sound, $ c_s\,{\sim}\,\lambda/\tau $ (where $ \lambda $~is
the free-path length, and $ \tau $~is the free-path time of the particles).
Then, the characteristic time of pressure relaxation in the region of
size~$ \Delta l $ will be
\begin{equation}
({\Delta}t)_p \approx \Delta l / c_s \sim (\Delta l / \lambda) \, \tau .
\label{eq:Relax_pressure}
\end{equation}
On the other hand, evolution of the density and temperature are described by
the equations of parabolic type:
$ \partial n / \partial t = - {\rm div} (k_n \nabla n) $ and
$ \partial T / \partial t = - {\rm div} (k_T \nabla T) $,
where $ k_{n,T}\,{\sim}\,{\lambda}^2/\tau $.
So, the characteristic relaxation times for the density and temperature
can be estimated as
\begin{equation}
({\Delta}t)_{n,T} \approx (\Delta l)^2 / k_{n,T} \sim
  (\Delta l / \lambda)^2 \tau .
\label{eq:Relax_dens_temp}
\end{equation}
Since $ \Delta l / \lambda \gg 1 $, then
$ ({\Delta}t)_{n,T} \gg ({\Delta}t)_p $, \textit{i.e.}, the equilibrium
value of pressure is quickly established throughout the system, while
fluctuations of the density and temperature relax at a much longer time scale.
\item
If an approximately constant pressure is established in the system, then
it follows immediately from the equation of state of the ideal gas,
$ p\,{=}\,nkT $ (where $ k $~is the Boltzmann constant), that the long-lived
fluctuations of the density~$ {\Delta}n $ and temperature~$ {\Delta}T $
should be anticorrelated: the smaller is the density, the greater is
the temperature, and \textit{vice versa}.
\end{enumerate}

\section{Application to the coronal plasmas}
\label{sec:Corona}

It was suggested by \citet{tel15} that just the above-mentioned
anticorrelation between the temperature and density developed during
the relaxation of strongly non-equilibrium plasmas can explain the observed
temperature profile in the solar corona, where temperature sharply increases
with decreasing density.
In general, this idea is well in agreement with the modern paradigm that
the corona is very irregular in the horizontal direction
\citep[\textit{e.g.},][]{gol10}, and the observed distribution of its
parameters is just the average outcome of many nonstationary processes
occurring in the individual magnetic flux tubes, as conjectured by
\citet{asc07}.

\begin{figure}
\begin{center}
\includegraphics[width=0.82\hsize]{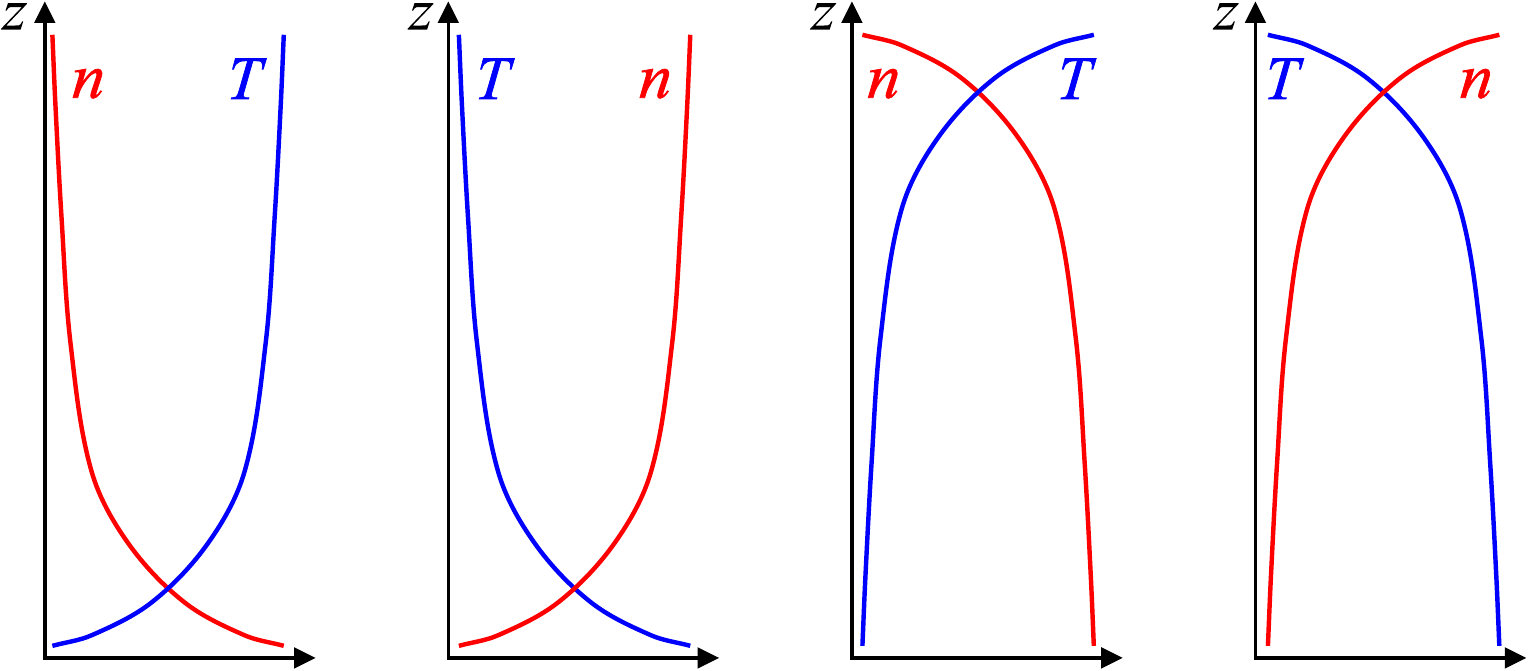}
\end{center}
\caption{Sketch of the various hypothetical profiles of the density~$ n(z) $
and temperature $ T(z) $ that should be formed in various parts of the solar
corona if they were caused by the spontaneous relaxation of the strongly
non-equilibrium plasma states.}
\label{fig:Profiles}
\end{figure}

Unfortunately, a closer inspection of the ``anticorrelation scenario''
shows that it can hardly serve as a viable realization of this paradigm.
There are two major obstacles already at the qualitative level (\textit{i.e.},
even before any quantitative estimates):
\begin{enumerate}
\item
Emergence of the anticorrelated temperature and density distributions
in the spin-type models assumes that both the temperature and density
gradients are formed spontaneously in a self-consistent way.
On the other hand, a strong density gradient in the solar corona evidently
results from the gravitational attraction to the center of the Sun
rather than from any plasma processes (associated with the Coulomb's
interaction).
So, one of the major prerequisites for the scenario discussed in
Sec.~\ref{sec:Spin-type} is not satisfied in the quiet solar corona,
as was already mentioned earlier in our paper \citep{dum16}.
\item
Moreover, even if the anticorrelated temperature and density profiles
would be self-consistently formed along the particular magnetic flux tubes,
the corresponding gradients should be directed randomly, as illustrated in
Fig.~\ref{fig:Profiles}.
So, the average distributions of the density and temperature with height
would be either constant (if spatial resolution of the observations is
insufficient to resolve the individual flux tubes) or oscillating in
the horizontal direction (if the resolution is sufficiently high).
Both these options are evidently irrelevant to the observable solar corona.
\end{enumerate}

\begin{figure}
\begin{center}
\includegraphics[width=0.62\hsize]{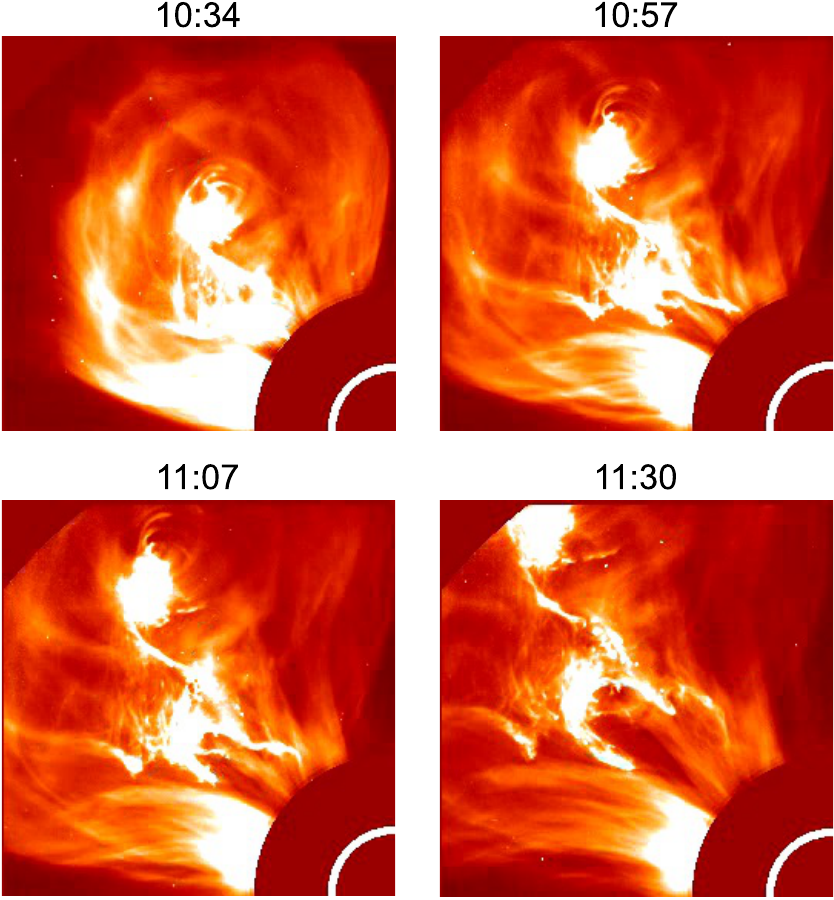}
\end{center}
\caption{Example of the powerful CME with a complex pattern of
inhomogeneities, observed on 04 January 2002 by LASCO/C2 instrument
onboard \textit{SOHO} satellite.
The corresponding coronogram was taken from the SOHO LASCO CME Catalog
\texttt{https://cdaw.gsfc.nasa.gov/CME{\_}list/};
courtesy of SOHO/LASCO consortium (ESA \& NASA).}
\label{fig:CME}
\end{figure}

Nevertheless, from our point of view, the ``anticorrelation scenario'' may
be still interesting for the solar physics, for example, if we consider
the coronal mass ejections (CME).
Really, both the above-mentioned obstacles will not take place for CMEs:
\begin{enumerate}
\item
Since plasmas ejected by CMEs move more or less freely, the gravitational
field is no longer of crucial importance in the local reference frame
of the substance (\textit{i.e.}, the situation is similar to
the microgravity conditions inside a spacecraft).
So, in the first approximation, the relaxation of temperature and density can
proceed in a self-consistent way, as required by the spin-type models.
\item
The distribution of plasma density in some CMEs exhibits a considerable
irregularity both in the vertical and horizontal direction; a particular
example is shown in Fig.~\ref{fig:CME}.
This reminds a set of the unusual height profiles represented in
Fig.~\ref{fig:Profiles}, because the density exhibits diverse nonmonotonic
variations along the various radii.
\end{enumerate}

\begin{figure}
\begin{center}
\includegraphics[width=0.55\hsize]{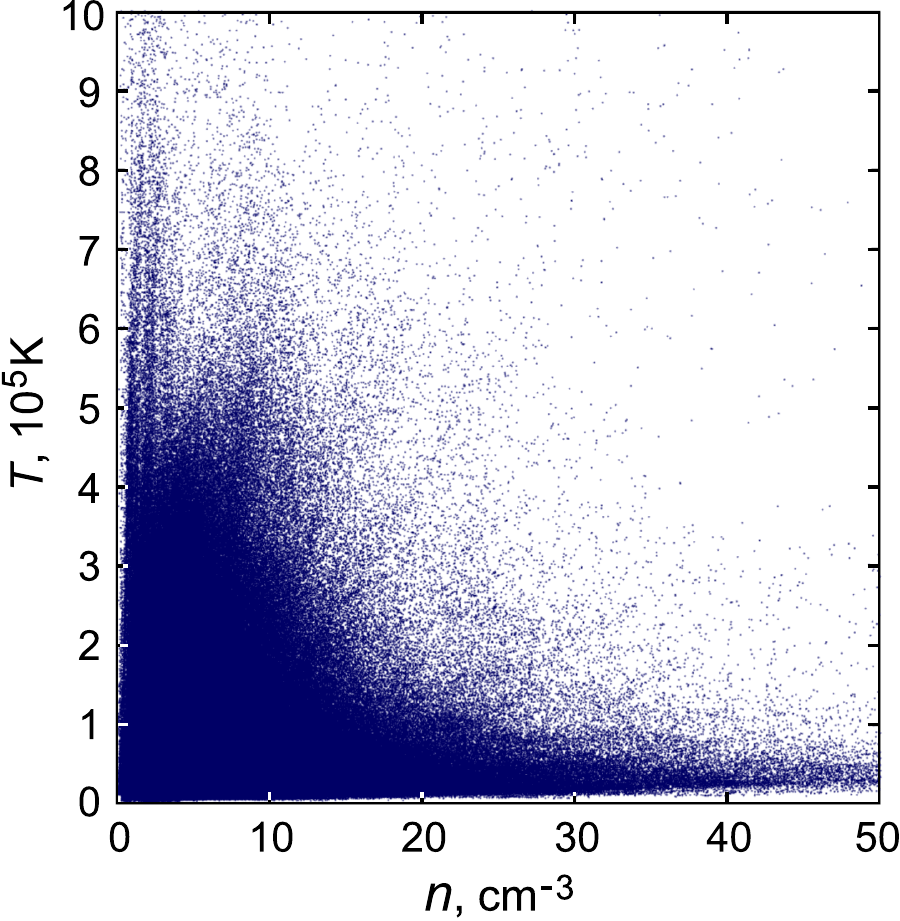}
\end{center}
\caption{Correlation diagram of the temperature and density in the solar
wing for the period from January 2009 to December 2017; courteously
prepared by A.T.~Lukashenko.}
\label{fig:Solar_wind}
\end{figure}

Unfortunately, it is impossible at the present time to perform a detailed
comparison of the density and temperature profiles in CMEs, because
the available coronograms reflect mostly the distribution of the electron
density, while there are no reliable measurements of both electron and ion
temperature inside CMEs \citep{gre18}.
However, one can expect that such measurements will be done in future
(\textit{e.g.}, by the methods of radioastronomy), and then searching for
the above-mentioned anticorrelations will be an interesting observational
task.

Yet another possible way for testing the anticorrelated distributions is
to analyze parameters of the solar wind near the Earth or in the Lagrange
point~L1, since the solar-wind plasma should keep imprints of its properties
at the closer distance to the Sun.
The correlation diagram of the temperature and density in the solar wind is
shown in Fig.~\ref{fig:Solar_wind}; the data were taken from the OMNIWeb site
(\texttt{https://omniweb.gsfc.nasa.gov}).

One can see that the entire diagram has a ``hyperbolic'' shape, which
indicates clearly to the anticorrelation of temperature and density.
Besides, there are well-expressed ``tails'' in the vertical and horizontal
direction.
It can be conjectured that they are associated just with the solar-wind
flows from CMEs.
(But, in principle, these features could be produced also during the
propagation of the solar wind to the Earth.)
The correlation coefficient between the temperature and density for the entire
above-mentioned period equals~$ -0.13 $, and its magnitude can be almost two
times greater for the narrower intervals (\textit{e.g.}, $ -0.23 $ for the
period from January 2009 to December 2010; and $ -0.19 $, from January 2016
to December 2017).
For more details on the correlation properties of the solar wind, see paper
by \citet{ves18} and references therein.

\section{Conclusions}
\label{sec:Concl}

\begin{enumerate}
\item
A universal property of relaxation of the strongly non-equilibrium plasmas
is a formation of the long-lived anticorrelated temperature and density
profiles.
This can be derived both from the mathematical formalism of the spin-type
Hamiltonians as well as from the more pictorial arguments presented in
Sec.~\ref{sec:Physical}.
\item
Unfortunately, as follows from the simple qualitative arguments, this
mechanism can hardly be applicable to explanation of the average temperature
profile in the quiet solar corona.
\item
However, the same processes may be important for other solar phenomena,
\textit{e.g.}, formation of the anticorrelated profiles in CMEs.
This should be an interesting observational task for the future research.
\item
The available measurements of the solar wind confirm the anticorrelation
between the temperature and density on large time intervals.
This may be either the contribution by CMEs or formed during the solar wind
propagation to the Earth.
\end{enumerate}

\section*{Acknowledgements}

I am grateful to S.I.~Bezrodnykh for valuable comments on the mathematical
issues, B.P.~Filippov and I.F.~Nikulin for the discussion of observational
data on CME, A.T.~Lukashenko for help in processing the data on the solar
wind, and especially to the anonymous referee for the suggestion to use
the data on the solar wind for testing the anticorrelated distributions.
I am also grateful to the Max Planck Institute for the Physics of Complex
Systems (Dresden, Germany) and, in particular, to its director J.-M.~Rost
for the support of my participation in a few conferences organized by this
institution.

The OMNI data were obtained from the GSFC/SPDF OMNIWeb interface at
\texttt{https://omniweb.gsfc.nasa.gov}.
The CME catalog used in the present work is generated and maintained at
the CDAW Data Center by NASA and The Catholic University of America in
cooperation with the Naval Research Laboratory.
SOHO is a project of international cooperation between ESA and NASA.


\end{document}